\documentclass[aps,prc,amsmath,amssymb,showpacs,superscriptaddress,twocolumn,footinbib,floatfix]{revtex4}
\usepackage{longtable,graphicx,dcolumn,epsfig,enumerate}
\begin{document}
\title{Photoactivation experiment on $^{197}$Au and its implications for the dipole strength in heavy nuclei}
\author{C. Nair}
\affiliation{Institut f\"ur Strahlenphysik, Forschungszentrum
             Dresden-Rossendorf, D-01314 Dresden, Germany}
\author{M. Erhard}
\affiliation{Institut f\"ur Strahlenphysik, Forschungszentrum
             Dresden-Rossendorf, D-01314 Dresden, Germany}
\author{A. R. Junghans}
\affiliation{Institut f\"ur Strahlenphysik, Forschungszentrum
             Dresden-Rossendorf, D-01314 Dresden, Germany}
\author{D. Bemmerer}
\affiliation{Institut f\"ur Strahlenphysik, Forschungszentrum
             Dresden-Rossendorf, D-01314 Dresden, Germany}
\author{R. Beyer}
\affiliation{Institut f\"ur Strahlenphysik, Forschungszentrum
             Dresden-Rossendorf, D-01314 Dresden, Germany}
\author{E. Grosse}
\affiliation{Institut f\"ur Strahlenphysik, Forschungszentrum
             Dresden-Rossendorf, D-01314 Dresden, Germany}
\affiliation{Institut f\"ur Kern- und Teilchenphysik,
             Technische Universit\"at Dresden, D-01062 Dresden, Germany}
\author{J. Klug}
\altaffiliation{Present address: Ringhals Nuclear Power
Plant, SE-43022 V\"ar\"obacka, Sweden} \affiliation{Institut f\"ur Strahlenphysik,
Forschungszentrum Dresden-Rossendorf, D-01314 Dresden, Germany}

\author{K. Kosev}
\affiliation{Institut f\"ur Strahlenphysik, Forschungszentrum
             Dresden-Rossendorf, D-01314 Dresden, Germany}
\author{G. Rusev}
\altaffiliation{Present address: Department of Physics, Duke
University, Triangle Universities Nuclear Laboratory, Durham, NC
27708, USA.} \affiliation{Institut f\"ur Strahlenphysik,
Forschungszentrum Dresden-Rossendorf, D-01314 Dresden, Germany}

\author{K. D. Schilling}
\affiliation{Institut f\"ur Strahlenphysik, Forschungszentrum
             Dresden-Rossendorf, D-01314 Dresden, Germany}
\author{R. Schwengner}
\affiliation{Institut f\"ur Strahlenphysik, Forschungszentrum
             Dresden-Rossendorf, D-01314 Dresden, Germany}
\author{A. Wagner}
\affiliation{Institut f\"ur Strahlenphysik, Forschungszentrum
             Dresden-Rossendorf, D-01314 Dresden, Germany}

\date{\today}

\begin{abstract}
The $^{197}$Au$(\gamma,n)$ reaction is used as an activation
standard for photodisintegration studies on astrophysically relevant
nuclei. At the bremsstrahlung facility of the superconducting
electron accelerator ELBE of FZ Dresden-Rossendorf, photoactivation
measurements on $^{197}$Au have been performed with bremsstrahlung
endpoint energies from 8.0 to 15.5 MeV. The measured activation
yield is compared with previous experiments as well as calculations
using Hauser-Feshbach statistical models. It is shown that the
experimental data are best described by a two-lorentzian
parametrization with taking the axial deformation of $^{197}$Au into
account. The experimental $^{197}$Au$(\gamma,n)$ reaction yield
measured at ELBE via the photoactivation method is found to be
consistent with previous experimental data using photon scattering
or neutron detection methods.
\end{abstract}
\maketitle

\section{Introduction}
\label{intro}Photonuclear processes are among the first nuclear
reactions ever studied in the laboratory~\cite{Bothe1939}. They have
provided important information about the giant dipole resonance
(GDR)~\cite{Dietrich1989} and play a vital role in our understanding
of the cosmic nucleosynthesis pointed out by Burbidge et
al.~\cite{B2FH}. In high temperature cosmic scenarios like exploding
supernovae, the photon flux is intense enough to cause the
photodisintegration of previously formed heavy nuclides. The
photonuclear cross sections are of importance for the understanding
of neutron capture in hot and neutron rich stellar environments,
where nuclei are likely to be excited from their ground states and
may simultaneously undergo capture. The usual laboratory study of
radiative neutron capture does not yield direct information on such
processes, but their inverse, photon induced neutron emission to
excited states may reveal respective information via the detailed
balance principle~\cite{Sonnabend2003}.

More generally, the combined information from photodisintegration
and photon scattering allows to derive the photon strength function
(PSF) below and above the separation energies. The PSF is an
essential ingredient for the modeling of astrophysical reaction
rates for network calculations of the cosmic nucleosynthesis. The
other component of such investigations is the Hauser-Feshbach
statistical model (HFM). Accurate experimental studies of the
excitation functions of photon induced processes allow sensitive
tests of the parameters entering the model calculations, - e.g.,
optical-model potentials, level densities and transmission
coefficients.

From photoneutron studies concentrating on the GDR region, the
accuracy needed for a detailed prediction of the yields of heavy
nuclei produced by neutron capture via s- and r-processes cannot be
reached~\cite{Kaeppeler2006}. For the neutron-deficient p-nuclei,
there is practically no experimental data existing in the
astrophysically relevant energy region~\cite{Arnould2003}. In view
of the emerging novel observations of isotopic yields in stellar
plasma and in gathered cosmic material, high accuracy network
calculations are of increasing interest.

The photoneutron cross section of $^{197}$Au has been measured by
various methods. It has been shown that the cross sections in the
isovector GDR region as measured at different laboratories may
differ beyond their statistical and systematic
uncertainties~\cite{Dietrich1989}. Recently, the photoneutron cross
section of $^{197}$Au has been measured with laser-induced Compton
backscattered (LC) photons at the TERAS storage ring at AIST
Tsukuba, Japan~\cite{Utsunomiya}. Photoactivation of Au has also
been investigated recently with bremsstrahlung at an extremely
stable clinical accelerator~\cite{Mohr2007}.

The $^{197}$Au$(\gamma,n)$ reaction is used as an activation
standard for photodisintegration studies on astrophysically relevant
nuclei. In this paper we present a study of the
$^{197}$Au$(\gamma,n)$ reaction for the whole region from the
neutron threshold $S_\mathrm{n}$ to beyond the top of the GDR with
an accuracy of nearly 10\%. The bremsstrahlung endpoint energies for
the measurements range from 8.0 to 15.5 MeV. Special care was taken
at each accelerator setting to measure the bremsstrahlung endpoint
energy without relying on the magnetic beam transport elements. The
photon flux was determined by an independent observation of photon
scattering from $^{11}$B exposed to the same photons as the Au
samples. The residual nucleus $^{196}$Au produced from the
$^{197}$Au$(\gamma,n)$ reaction was studied by $\gamma$-ray
spectroscopy.

Secs.~\ref{sect:setup} and~\ref{sect:datanal} describe the
experimental procedure and the data analysis. In
Sec.~\ref{sect:previousexp}, the experimental activation yield is
compared to the yield calculated using cross sections from previous
experiments on $^{197}$Au$(\gamma,n)$.

In Sec.~\ref{sect:statisticalmodel}, the experimental yield is
compared with Hauser-Feshbach model calculations. It is shown that
the predictions of these models deviate from the measured activation
yield.

A phenomenological parametrization of the photon strength function
is proposed which describes the experimental data and extrapolates
it well to the threshold region. Sec.~\ref{sect:parametrization} of
this paper is devoted to the description of this parametrization and
to its comparison with data from photoneutron and photon-scattering
studies as well as the comparison with other descriptions of the
photon strength function.

\section{Experimental Setup}\label{sect:setup}
The experiments were performed at the superconducting electron
accelerator ELBE (Electron Linear accelerator of high Brilliance and
low Emittance) of the Forschungszentrum Dresden-Rossendorf. ELBE can
produce intense bremsstrahlung beams with endpoint energies from 6
to 18 MeV. With these beam parameters both photon scattering and
photodisintegration reactions have been
measured~\cite{Rusev2006,Rusev2006-EPJ,Schwengner2007,Wagner2008,Rusev2008}.
\begin{figure}
\begin{center}
\includegraphics[width=8cm,angle=0]{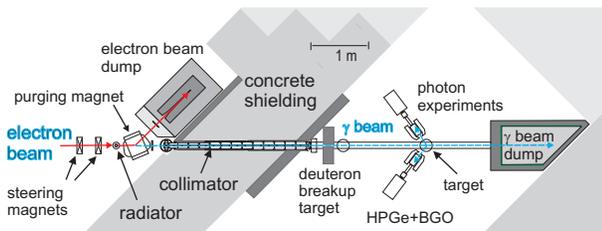}
\end{center}
\caption{\label{fig:setup}(Color online) The bremsstrahlung facility
at ELBE. The Au targets were irradiated together with $^{11}$B
samples at the target site. The photons scattered from $^{11}$B
samples were measured using four 100\% HPGe detectors with BGO
escape-suppression shields, two of which were mounted vertically
(not shown). The endpoint energy of the bremsstrahlung was
determined from the proton spectrum of the deuteron breakup
reaction.}
\end{figure}

The bremsstrahlung facility is shown in Fig.~\ref{fig:setup}. The
electron beam is focused onto a niobium radiator with thicknesses
varying between 1.7~mg/cm$^2$ and 10~mg/cm$^2$ (corresponding to
$1.6\cdot10^{-4}$ and $1\cdot10^{-3}$ radiation lengths) which
creates typical "thin target" bremsstrahlung. After passing the
radiator, the electrons are deflected by a dipole magnet and dumped
to a graphite cylinder mounted on insulating rods surrounded by a
water cooled vacuum vessel (electron beam dump, see figure). A
collimator placed 1 m behind the radiator is used to form a beam
with defined diameter out of the spatial distribution of photons.
The collimator is made from high-purity aluminum and is fixed within
the 1.6 m thick wall of heavy concrete between the accelerator hall
and the experimental cave. An aluminum cylinder of 10 cm diameter
and 10 cm length placed in a vacuum chamber in front of the entrance
of the collimator acts as a hardener which absorbs mainly low energy
photons and  thus 'hardens' the photon spectrum.

At the target site, the bremsstrahlung beam is collimated onto the
$^{197}$Au targets sandwiched with a $^{11}$B sample. The photon
flux is determined experimentally by means of the known integrated
cross sections of the states in $^{11}$B depopulating via $\gamma$
rays. Photons scattered from $^{11}$B are measured with four
high-purity germanium (HPGe) detectors of 100\% relative efficiency
which are surrounded by escape-suppression shields consisting of
bismuth-germanate (BGO) scintillation detectors. The experimental
procedure has been described in detail
elsewhere~\cite{Schwengner2005,Wagner2005}.

The $^{197}$Au targets used were thin discs with a typical mass of
about 200 mg, thickness 0.02 mm, and diameter 20 mm. The number of
activated nuclei produced during the activation was determined
offline by measuring the decay of daughter nuclei in a low-level
counting setup by HPGe detectors with relative efficiencies of 90\%
or 60\%.

The endpoint energy of the bremsstrahlung distribution is determined
by measuring protons from the photodisintegration of the deuteron
(see Fig.~\ref{fig:setup}, deuteron breakup target) with silicon
detectors. From the maximum energy of the emitted protons, the
maximum energy of the incident photons can be deduced. This is
described in detail in Sec.~\ref{sect:deutronbreakup}. During the
experiment, energy drifts of the electron linac have been kept to
below 1\% using non-destructive beam-diagnostics of the transverse
beam dispersion and an active beam-stabilization control loop.

\section{Data Analysis}\label{sect:datanal}
The data deduction and analysis methods will be described in detail
in the following sections. The discussion is split into three
parts:
\begin{enumerate}[(A)]
\item Decays observable following the $^{197}$Au$(\gamma,n)$
reaction and determination of the photoactivation yield;
\item Experimental determination of the photon flux at the scattering
site; and
\item Bremsstrahlung endpoint energy determination using
the deuteron breakup reaction.
\end{enumerate}
\subsection{$^{196}$Au decay}\label{sect:yield}
\begin{table}
\caption{ Decay properties of the $^{196}$Au nucleus.}
\label{tb:decayproperties}
\begin{ruledtabular}
\begin{tabular}{ccc}
Nuclide\footnotemark[1] & $E_{\gamma}$ (keV)\footnotemark[2] &
p\footnotemark[3]  \\
\hline $^{196}$Pt
    & 333.03(5) & 0.229(10)\\
$^{196}$Pt
    & 355.73(5) & 0.87(3)\\
$^{196}$Hg
    & 426.10(8) & 0.066(3)\\
\hline
\end{tabular}
\end{ruledtabular}
\footnotetext[1]{Daughter nuclide from $^{196}$Au decay}
\footnotetext[2]{
 Energy of the transition with absolute uncertainty given in parentheses. }
\footnotetext[3]{
 Photon emission probability per decay with absolute uncertainty given in parentheses. }
\end{table}
\begin{figure}
\begin{center}
\includegraphics[width=6.2 cm,angle=270]{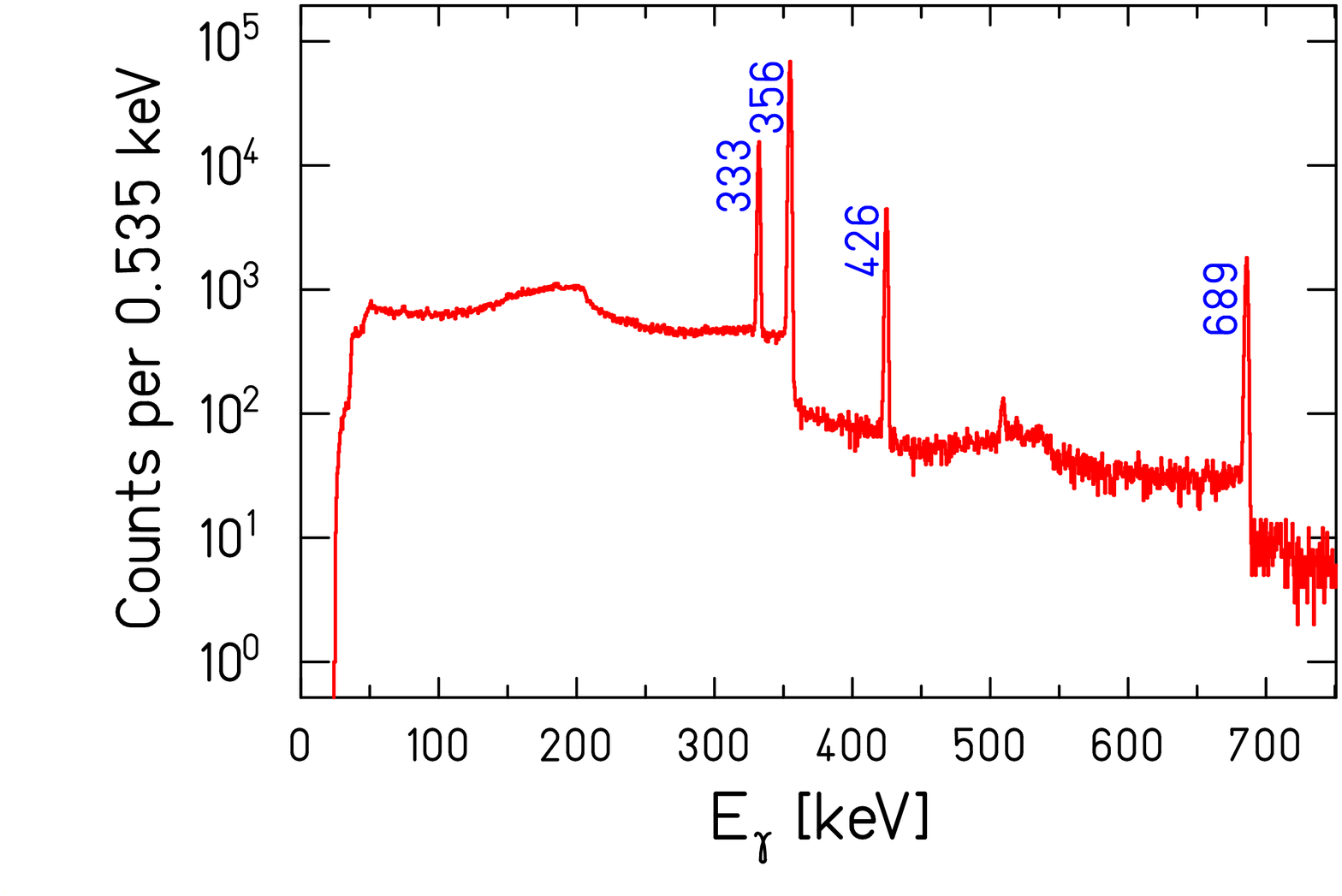}
\end{center}
\caption{\label{fig:196Audecay}(Color online) Spectrum of an
irradiated $^{197}$Au target. The target was placed on the top of a
HPGe detector with 90\% relative efficiency. The peaks originating
from the $^{196}$Au decay are marked. The $\gamma$-line at 689 keV
is the sum of the $\gamma$-transitions with energies 333 and 356
keV. A 1.5 mm thick Cd absorber was used to suppress coincidence
summing and low energy X-rays.}
\end{figure}

The $^{197}$Au$(\gamma,n)$ reaction produces the unstable nucleus
$^{196}$Au which decays either to $^{196}$Pt by electron capture or
positron emission (EC+$\beta^{+}$) or to $^{196}$Hg by beta-decay
($\beta^{-}$). A typical decay spectrum of a $^{197}$Au sample
irradiated with a bremsstrahlung endpoint energy of 14.5 MeV for 17
hours is given in Fig.~\ref{fig:196Audecay}. The prominent peaks in
the decay of $^{196}$Au used for analysis are marked in
Fig.~\ref{fig:196Audecay} and are given in
Table~\ref{tb:decayproperties}. The decay properties given in the
table are adopted from Ref.~\cite{NNDC}.

The absolute photopeak efficiency of the counting setup has been
determined with several calibration sources from PTB and Amersham
(systematic uncertainty in activity 0.6-1.5\%) in the energy range
from 0.12 to 1.9 MeV~\footnotemark[1]\footnotetext[1]{PTB:
Physikalisch-Technische Bundesanstalt, Fachbereich 6.1, Bundesallee
100, Braunschweig, Germany;\newline Amersham: ISOTRAK AEA Technology
QSA, Gieselweg 1, Braunschweig, Germany.}. The absolute efficiency
was simulated for a realistic geometry using the Monte Carlo code
GEANT3~\cite{GEANT3} and was fitted to the measured data.
Coincidence summing effects depend strongly on the decay scheme.
They were determined very precisely for the corresponding counting
geometry. The distance between the surface of the endcap and
detector crystal was cross-checked by X-ray radiography. The number
of $\gamma$-rays counted in the peaks at 333 and 356 keV have been
corrected for 'summing-out' events using the method described in
Ref.~\cite{Debertin}. For the transition at 333 keV, the coincidence
summing correction amounts to 24\% and for 356 keV it is 6\%, both
with a relative uncertainty of 5\%.

In a photoactivation experiment, the number of radioactive nuclei
$N_{\mathrm{act}}(E_0)$ produced is proportional to the integral of
the absolute photon fluence $\Phi_{\gamma}(E,E_0)$ times the
photodisintegration cross section $\sigma _{\gamma ,\mathrm{n}}(E)$
integrated from the reaction threshold energy $E_{\mathrm{thr}}$ up
to the endpoint energy $E_0$ of the bremsstrahlung spectrum.
\begin{equation}
N_{\mathrm{act}}(E_0) =  N_{\mathrm{tar}} \cdot
\int_{E_{\mathrm{thr}}}^{E_{\mathrm{0}}} \sigma_{\mathrm{\gamma
,n}}(E)\cdot \Phi_{\gamma}(E,E_0)\,dE \label{eqn:actnuclei}
\end{equation}
The number of radioactive nuclei $N_{act}(E_0)$ is determined
experimentally by measuring the activity of the irradiated sample
using:
\begin{equation}
N_{\mathrm{act}}(E_0) =
\frac{N_{\gamma}(E_{\gamma},E_0)\cdot\kappa_{\mathrm{corr}}}{\varepsilon
(E_{\gamma})\cdot p(E_\gamma)} \label{eqn:yint}
\end{equation}
$ N_{\gamma}(E_{\gamma},E_0)$, $\varepsilon(E_{\gamma})$,
$p(E_\gamma)$ denote the dead-time and pile-up corrected full-energy
peak counts of the observed transition, the absolute efficiency of
the detector at the energy $E_{\gamma}$ and the emission probability
of the photon with energy $E_\gamma$, respectively.

The factor $\kappa_{\mathrm{corr}}$ in Eq.~(\ref{eqn:yint}) is given
by
\begin{equation}
\kappa_{\mathrm{corr}}
=\frac{\exp{(\frac{t_{\mathrm{loss}}}{\tau})}}{1-\exp{(\frac{-t_{\mathrm{meas}}}{\tau})}}\cdot
\frac{\frac{t_{\mathrm{irr}}}{\tau}}{1-\exp{(\frac{-t_{\mathrm{irr}}}{\tau})}}
\label{eqn:kappacorr}
\end{equation}
This expression determines the number of radioactive nuclei from
their decays measured during the time $t_{\mathrm{meas}}$. It also
takes into account decay losses during irradiation
($t_{\mathrm{irr}}$) and in between the end of the irradiation and
the beginning of the measurement ($t_{\mathrm{loss}}$). The mean
life time of the radioactive nucleus produced during the
photoactivation is denoted by $\tau$. The decay time constants of
$^{196}$Au and $^{198}$Au have been confirmed in a precision
measurement using targets produced in the scope of the present
experiment~\cite{Ruprecht2008}.

The activation yield is denoted by $Y_{\mathrm{act}}$ and is
expressed as the ratio of the number of activated nuclei to the
number of target atoms in the sample. For the $^{197}$Au$(\gamma,n)$
reaction,
\begin{equation}\label{eqn:yield}
 Y_{\mathrm{act}} = \frac{N_{\mathrm{act}}(^{196}\rm{Au})}{N_{\mathrm{tar}}(^{197}\rm{Au})}
 \end{equation}
Using Eq.~(\ref{eqn:actnuclei}), the activation yield can be
calculated from $\sigma_{\gamma ,n}(E)$ data with the known
bremsstrahlung spectrum. In this way measured activation yields can
be compared with the experimental or theoretical cross section data.

\subsection{The photon flux}
\label{sect:flux} In the present study, the photon flux was
determined from the elastic photon scattering from a $^{11}$B sample
sandwiched with the Au activation target. Four HPGe-detectors (two
at 90$^{\circ}$ and two at 127$^{\circ}$) were used for this
measurement (see Ref.~\cite{Schwengner2005} for details).The photon
fluence is determined experimentally using the formula:
\begin{eqnarray}
\label{eqn:flux} \Phi^{\rm
}_\gamma(\mathrm{E}_\gamma)=\frac{N_\gamma(\mathrm{E}_\gamma)}
 {\varepsilon(\mathrm{E}_\gamma)\cdot N_{\rm tar}
\cdot I_{\rm s}\cdot\mathrm{W}(\theta)}
\end{eqnarray}
$ N_{\gamma}(E_{\gamma})$, $\varepsilon(E_{\gamma})$, $N_{\rm tar}$
represents the dead-time and pile-up corrected full-energy peak
counts of the resonant transition, the absolute efficiency of the
detector at the energy $E_{\gamma}$, and the number of target atoms
in the $^{11}$B sample. $W(\theta)$ is the angular correlation
between the incoming and scattered photon and $I_{\rm s}$ denotes
the integrated scattering cross section.

The decay properties of calibration transitions were adopted from
the online library of Evaluated Nuclear Structure Data Files (ENSDF)
which refers to the revised Ajzenberg-Selove compilation (see Table
11.4, Ref.~\cite{Selove1990}). The absolute photopeak efficiency has
been determined with calibration sources for energies up to 1.9 MeV.
For extrapolating the efficiency to higher energies, a GEANT3
simulation under realistic geometry was used (see Fig.4,
Ref.~\cite{Rusev2008}). The simulations were normalized to the
measured efficiency at energies below 1.9 MeV.

In a typical case, the $^{11}$B target used was of metallic boron
powder with an enrichment of 99.5\%, mass areal density of 1.43 g
cm$^{-2}$, and an effective density of 1.6 g cm$^{-3}$. Energy
dependent nuclear self absorption corrections were applied using the
formalism given in Ref.~\cite{Skorka1975}. For example, for the
transition at 7.288 MeV, the nuclear self absorption correction
amounts to about 7.5\% when using a target with the specifications
given above.
\begin{figure}
\begin{center}
\includegraphics[width=6.2 cm,angle=270]{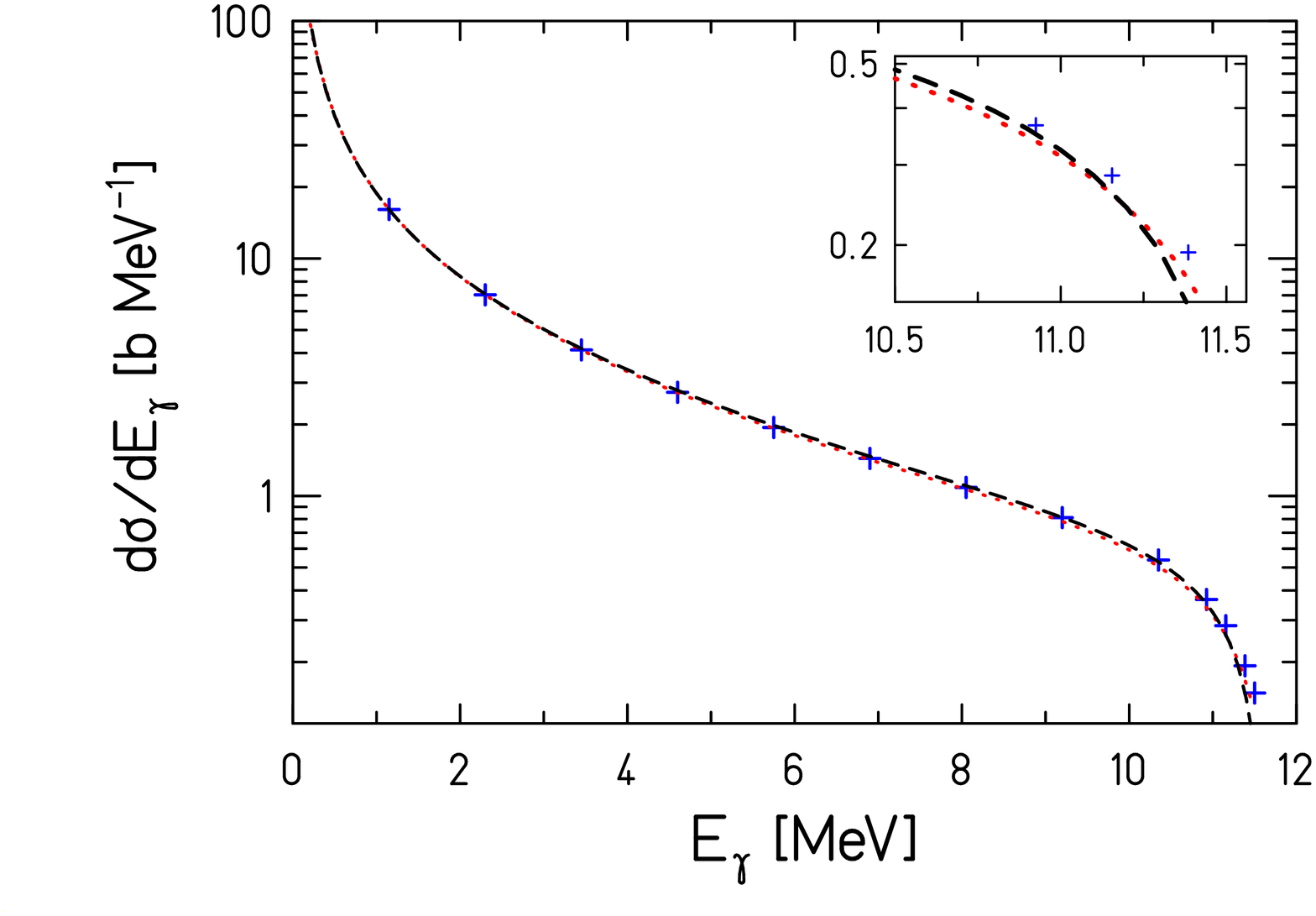}
\end{center}
\caption{\label{fig:brems-theory}(Color online) Comparison of
theoretical bremsstrahlung cross sections for the Nb radiator for an
incident electron endpoint energy of 11.5 MeV. Dashed and dotted
lines correspond to the bremsstrahlung distributions by
Schiff~\cite{Schiff1951} and Haug~\cite{Haug2008} whereas values
created from the Seltzer and Berger~\cite{Seltzer1986} tables are
shown as  symbols (+).}
\end{figure}
\begin{figure}
\begin{center}
\includegraphics[width=6.2 cm,angle=270]{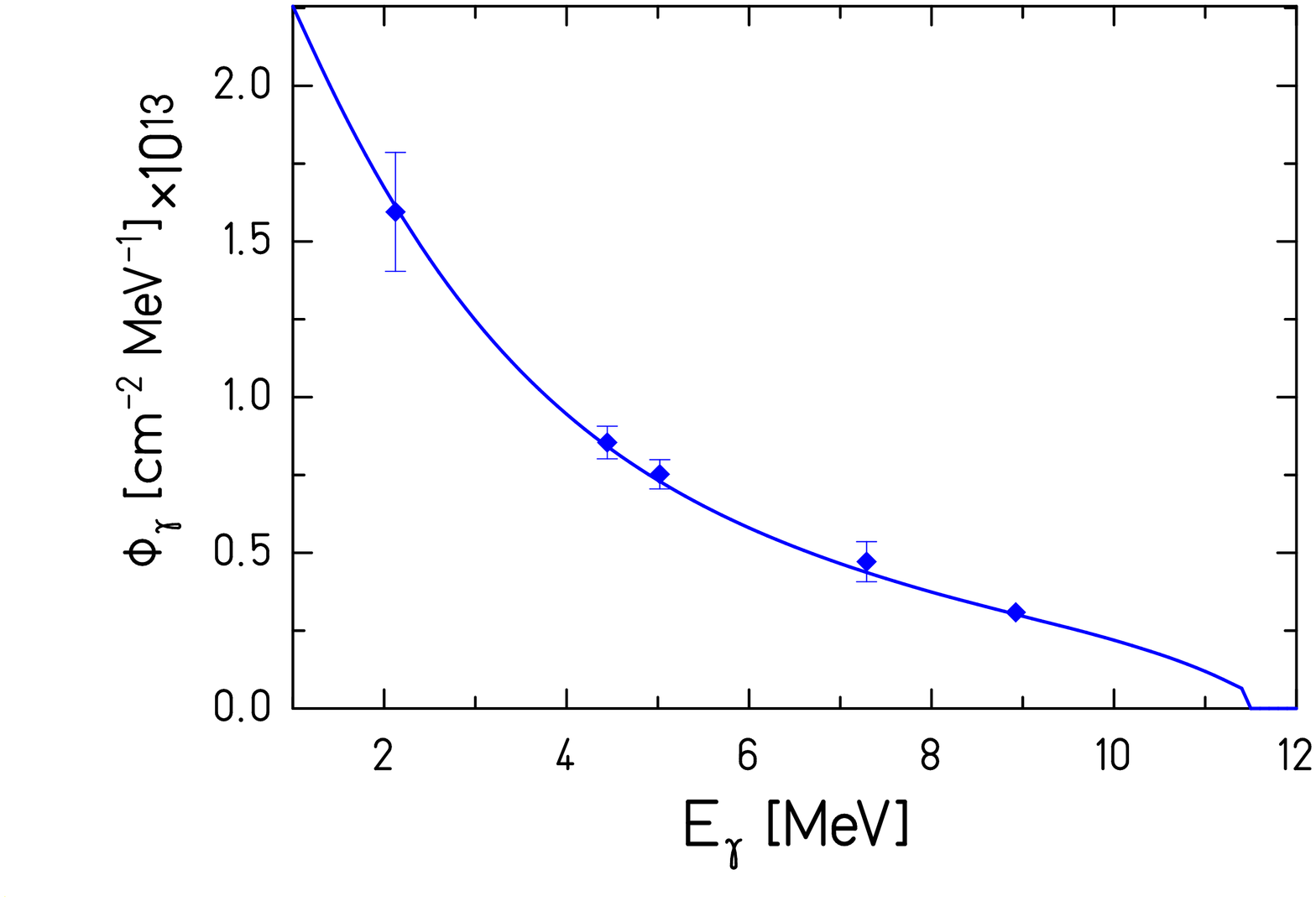}
\end{center}
\caption{\label{fig:brems-expt}(Color online) Absolute photon
fluence measured from the scattered photons in $^{11}$B is compared
with the Seltzer and Berger bremsstrahlung spectra with hardener
corrections. The fluence measured with different transitions in
$^{11}$B agree to the simulated curve to within 6\%.}
\end{figure}

The bremsstrahlung spectrum is well approximated by the theoretical
bremsstrahlung distribution for a thin niobium target. Different
approaches are compared for the niobium radiator for an incident
electron endpoint energy of 11.5 MeV as shown in
Fig.~\ref{fig:brems-theory}. They agree well with recent quantum
mechanical calculations by Haug~\cite{Haug2008,Roche1972} which use
the atomic shielding effects given in Ref.~\cite{Salvat1987}.

At the low-energy side of the spectrum the different theoretical
approaches are not distinguishable from each other and agree within
1 percent. Near endpoint, the theoretical models differ by about
20\% (see inset, Fig.~\ref{fig:brems-theory}). The theoretical
description of the high-energy end of the bremsstrahlung
distribution has a systematic effect on the calculation of the
activation yield from a given photoneutron cross section (see
section \ref{sect:previousexp}). For an end-point energy of 9 MeV or
higher, the  $^{197}$Au$(\gamma,n)$ activation yield calculated with
the cross section from Haug would be $5\%$ lower than that
calculated with the cross section from Seltzer and Berger. Below 9
MeV this effect increases up to $30 \%$.

The experimental photon fluence determined from the
$^{11}$B$(\gamma,\gamma${'}$)$ reaction for the $\gamma$-transitions
at 2.125, 4.446, 5.022, 7.288, and 8.924 MeV are shown in
Fig.~\ref{fig:brems-expt}. The bremsstrahlung spectrum was simulated
using MCNP~\cite{MCNP} to take into account the effects of the
aluminum hardener situated behind the niobium radiator. In MCNP, the
bremsstrahlung cross sections from Seltzer and Berger are used. The
simulated bremsstrahlung spectrum has been normalized to the
measured absolute photon fluence at the transition energies of
$^{11}$B. The systematic deviations between the simulated curve and
the experimental points are about 6\%.

In the fluence determination procedure discussed above, the
statistical contribution to the uncertainties from the
gamma-counting is quite small and is of the order of 0.5-2\%. The
systematic uncertainty in the extrapolation of efficiency is
estimated to be about 5\% in the energy range of the observed
transitions in $^{11}$B.

\subsection{Determination of bremsstrahlung endpoint energy}\label{sect:deutronbreakup}
For the experiments described here, it is necessary to measure the
endpoint energy of the bremsstrahlung spectra precisely. An on-line
measurement of the beam energy is attained using the dispersion
inside a dipole magnet with a magnetic field integral $\int B$dl
known to about 1\% only~\cite{Justus2007}. Therefore, we employed a
different method for the beam energy determination which is based on
the spectroscopy of protons in the photodisintegration of the
deuteron - the $\mathrm{^2H(\gamma,p)n}$ reaction. From the pure
two-body kinematics, the energy of the incident photon can be
deduced directly from the measured energy of the emitted proton.

\begin{figure}
\begin{center}
\includegraphics[width=6.2 cm,angle=270]{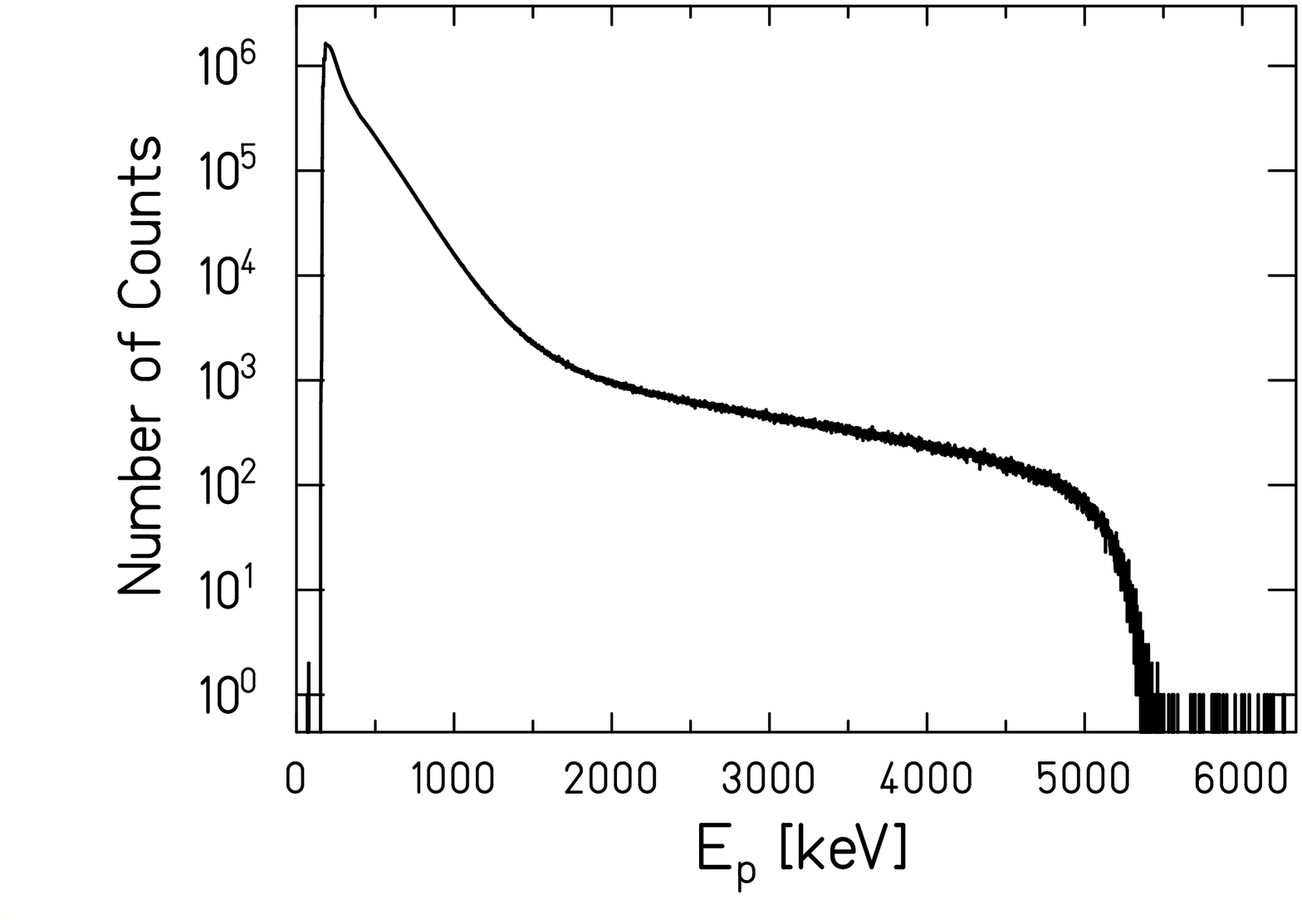}
\end{center}
\caption{\label{fig:Si-spectra}Proton spectrum from the
photodisintegration of deuterons, recorded with Si detectors of
500~$\mu$m thickness during an irradiation with incident electron
energy 13.2 MeV.}
\end{figure}

The protons from the photodisintegration of the deuteron are
detected by a setup of four silicon detectors (Ion-Implanted-Silicon
Charged-Particle Detectors, type ORTEC
ULTRA~\footnotemark[2]\footnotetext[2]{ORTEC, 801 South Illinois
Avenue, Oak Ridge, TN  37830, USA.}) placed at a distance of 115 mm
from the beam axis and at azimuthal angles of 0$^{\circ}$,
90$^{\circ}$, 180$^{\circ}$ and 270$^{\circ}$ with respect to the
photon beam. The detectors have a thickness of 500~$\mu$m and a
sensitive area of 600 mm$^2$. A 4 mg/cm$^2$ thick polyethylene film,
in which hydrogen is substituted by deuterium
(CD$_{2}$)~\footnotemark[3]\footnotetext[3]{Courtesy : D. K. Geiger,
SUNY Geneseo, NY 14454, USA.} is used as a target. The CD$_{2}$
target is positioned parallel to the incident beam such that its
surface is observed by all four detectors under 45$^{\circ}$. A
typical spectrum is shown in Fig.~\ref{fig:Si-spectra}. The
low-energy part of the spectrum below 2.5 MeV is not useful as it is
dominated by beam induced background.
\begin{figure}
\begin{center}
\includegraphics[width=6.2 cm,angle=270]{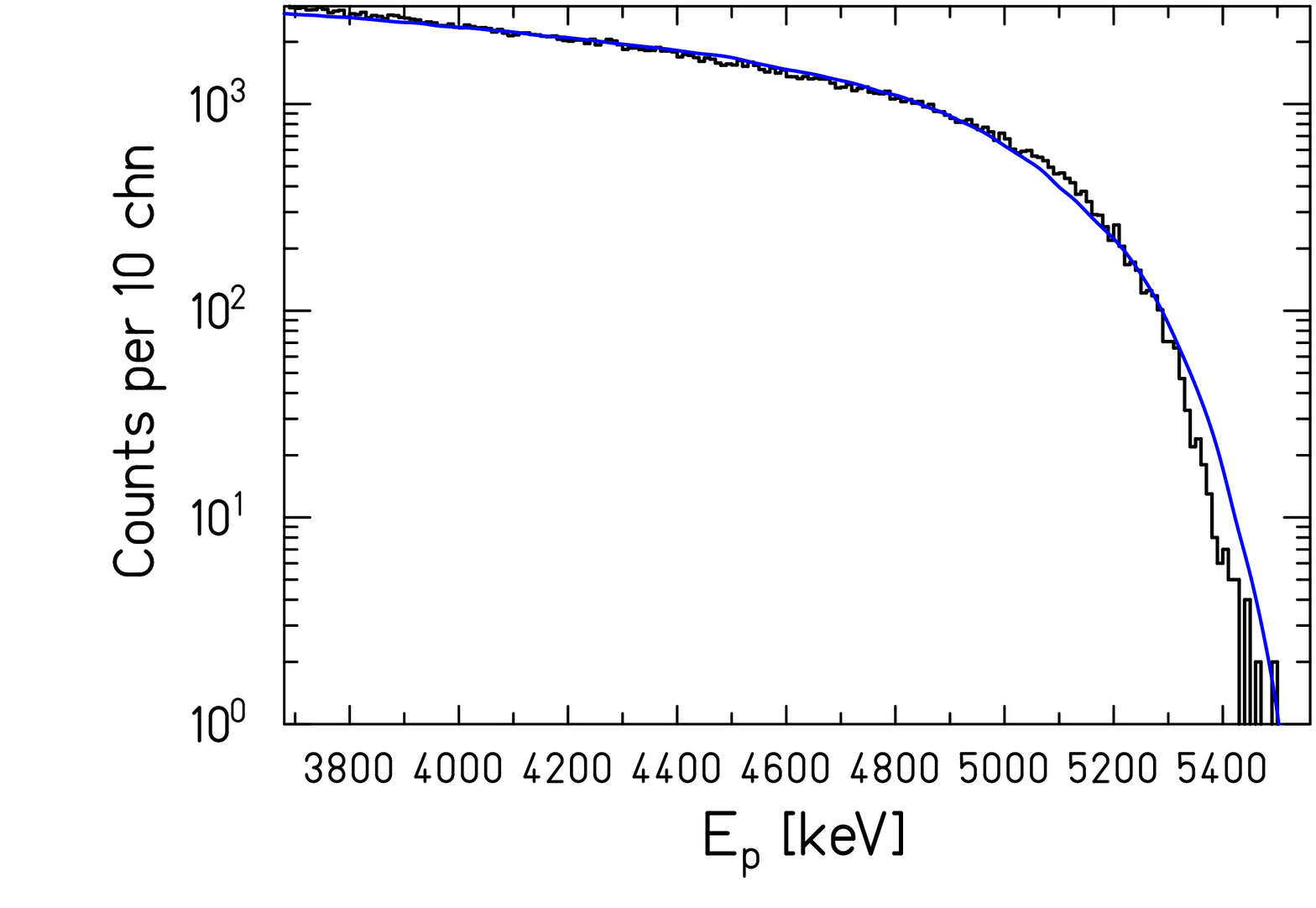}
\end{center}
\caption{\label{fig:Si-spectra-fit}(Color online) The simulated
proton spectrum (line) is fitted to the measured spectrum
(histogram) from Fig.~\ref{fig:Si-spectra}.}
\end{figure}

In order to determine the endpoint energy, a simulated spectrum is
fitted to the measured proton spectrum. The simulation takes into
account the deuteron breakup kinematics, geometry of the detector
setup, energy loss of the protons inside the CD$_{2}$ film as well
as the energy spread of the electron beam. The fit to the measured
spectra is shown in Fig.~\ref{fig:Si-spectra-fit}. The statistical
error from the fit amounts to 2-8 keV for the range of energies
described here. The systematic deviation of the experimental spectra
to the simulated one is 40 keV. This is inherent to all experiments
but significant only for endpoint energies close-above the
neutron-emission threshold of $^{197}$Au$(\gamma,n)$ reaction.
\section{Results and Discussion}
Photoactivation experiments with bremsstrahlung have the limitation
that the data need to be unfolded to obtain a cross
section~\cite{Berman1975}. This requires precise knowledge of the
bremsstrahlung spectrum especially close to the endpoint, and data
with very high counting statistics on a fine grid of endpoint
energies. In this work, the measured photoactivation yield is
presented and compared to calculated yield curves.

For the $^{197}$Au$(\gamma,n)$ reaction, the photoactivation yield
is determined as described by Eq.~(\ref{eqn:yield}). The activation
yield is normalized to the photon fluence for the corresponding
measurement as discussed in Sec.~\ref{sect:flux}. The endpoint
energies were determined from the photodisintegration of the
deuteron as explained in Sect.~\ref{sect:deutronbreakup}. The
experimental activation yield normalized to the photon fluence is
compared with previous experimental data as well as model
calculations.

\subsection{Activation Yield : Comparison with previous experiments}\label{sect:previousexp}
\begin{figure}
\begin{center}
\includegraphics[trim= 0 50 0 0 ,width=6 cm,angle=270]{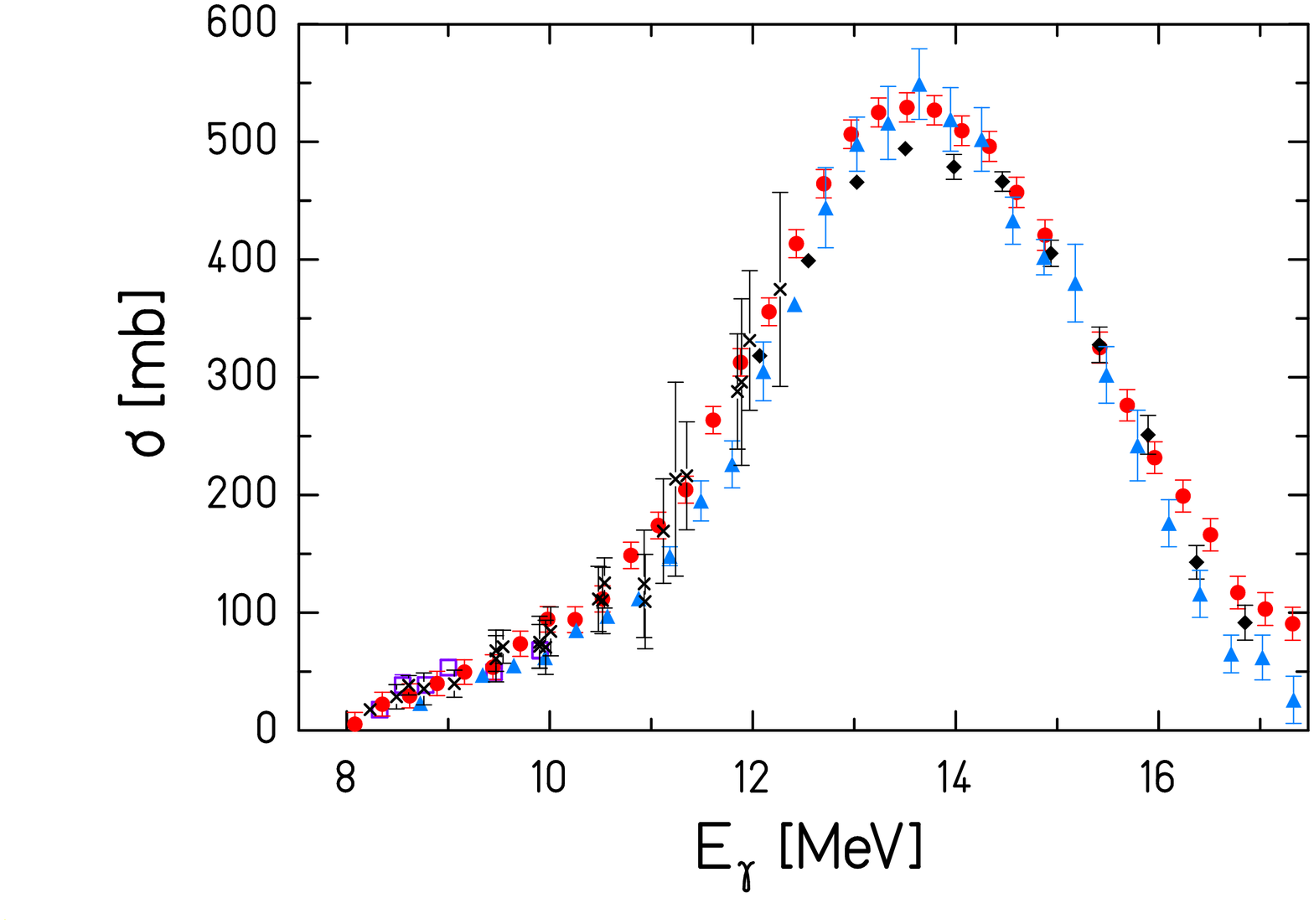}
\end{center}
\caption{\label{fig:crosssec}(Color online) Photoneutron cross
sections for $^{197}$Au$(\gamma,n)$ from previous experiments. The
symbols denote data from the respective experiments : triangles -
Fultz et al.~\cite{Fultz1962}, diamonds - Berman et
al.~\cite{Berman1987}, circles - Veyssiere et
al.~\cite{Veyssiere1970}. Some $^{197}$Au$(\gamma,n)$ cross section
data  below 10 MeV have been derived from bremsstrahlung activation
by Vogt et al. (open squares)~\cite{Vogt2002}. Also shown are cross
sections determined from Laser-Compton scattering by Hara et al.
($\times$) \cite{Utsunomiya}.}
\end{figure}

In this section, the activation yield from the ELBE experiments is
compared to calculated yields using cross sections measured in
previous experiments. A comparison of the $^{197}$Au$(\gamma,n)$
cross sections from previous
experiments~\cite{Fultz1962,Berman1987,Veyssiere1970,Vogt2002} is
given in Fig.~\ref{fig:crosssec}. At the Lawrence Livermore National
Laboratory (LLNL), the photoneutron cross section of the nucleus
$^{197}$Au has been measured with quasi-monoenergetic photons from
the positron annihilation technique. There are two sets of published
data - first by Fultz et al.~\cite{Fultz1962} and later by Berman et
al.~\cite{Berman1987}. The same technique has been used by Veyssiere
et al.~\cite{Veyssiere1970} at Saclay (France) for studying
photoneutron reactions on $^{197}$Au. The results from Livermore and
Saclay are not in agreement, revealing the differences in the
neutron multiplicity determination procedure used in both
laboratories.

Berman et al. have remeasured  photoneutron cross sections with
quasi-monoenergetic photons at LLNL, with special emphasis on
determining the absolute cross section at energies across the peak
of the GDR. Based on this experiment, Berman et al. resolves the
differences by recommending a 7\% scaling on the Veyssiere data and
ignoring the Fultz data(see Table VI, Ref.~\cite{Berman1987}). We
adopt this recommendation for comparing the ELBE data with the
previously reported values.

At the Laser-Compton scattering facility at the TERAS storage ring
at AIST Tsukuba, quasi-monoenergetic photons were used to study
photoneutrons from $^{197}$Au$(\gamma,n)$ up to 12.4 MeV. These data
agree very well with the data measured with the positron anihilation
technique but as a photon difference method was used they have a
rather large experimental uncertainty.

The photoneutron cross section of  $^{197}$Au for energies close
above the $(\gamma,n)$ threshold has been deduced by Vogt et
al.~\cite{Vogt2002} using photoactivation with bremsstrahlung at the
S-DALINAC (Darmstadt). The cross sections are in agreement with
Veyssiere et al., but exist only for endpoint energies between 8.0
MeV  and 10.0 MeV.

The total nuclear photoabsorption cross section of $^{197}$Au was
measured at the synchrotron facility of the Institute of Nuclear
Research (Moscow) by Gurevich et al.~\cite{Gurevich1981}. Even
though the data agree with the measurements by Veyssiere et al.,
they exhibit significant scatter (Fig. 2, Ref.~\cite{Gurevich1981}).
The tabulated errors are quite big and therefore were not included
for comparison with the ELBE data reported here. The photoneutron
yield for $^{197}$Au was measured by Sorokin et
al.~\cite{Sorokin1973}, at the Betatron (Moscow State University)
and the cross sections were deduced by the
Penfold-Leiss~\cite{Penfold1959} method. This experiment was done
with an energy resolution of 0.5 MeV for the range of energies
considered here. The results from Sorokin et al. are not included in
the present discussion because the uncertainties resulting from the
unfolding process are very large and the data differ significantly
from the previous experimental data.
\begin{figure}
\begin{center}
\includegraphics[width=6.2 cm,angle=270]{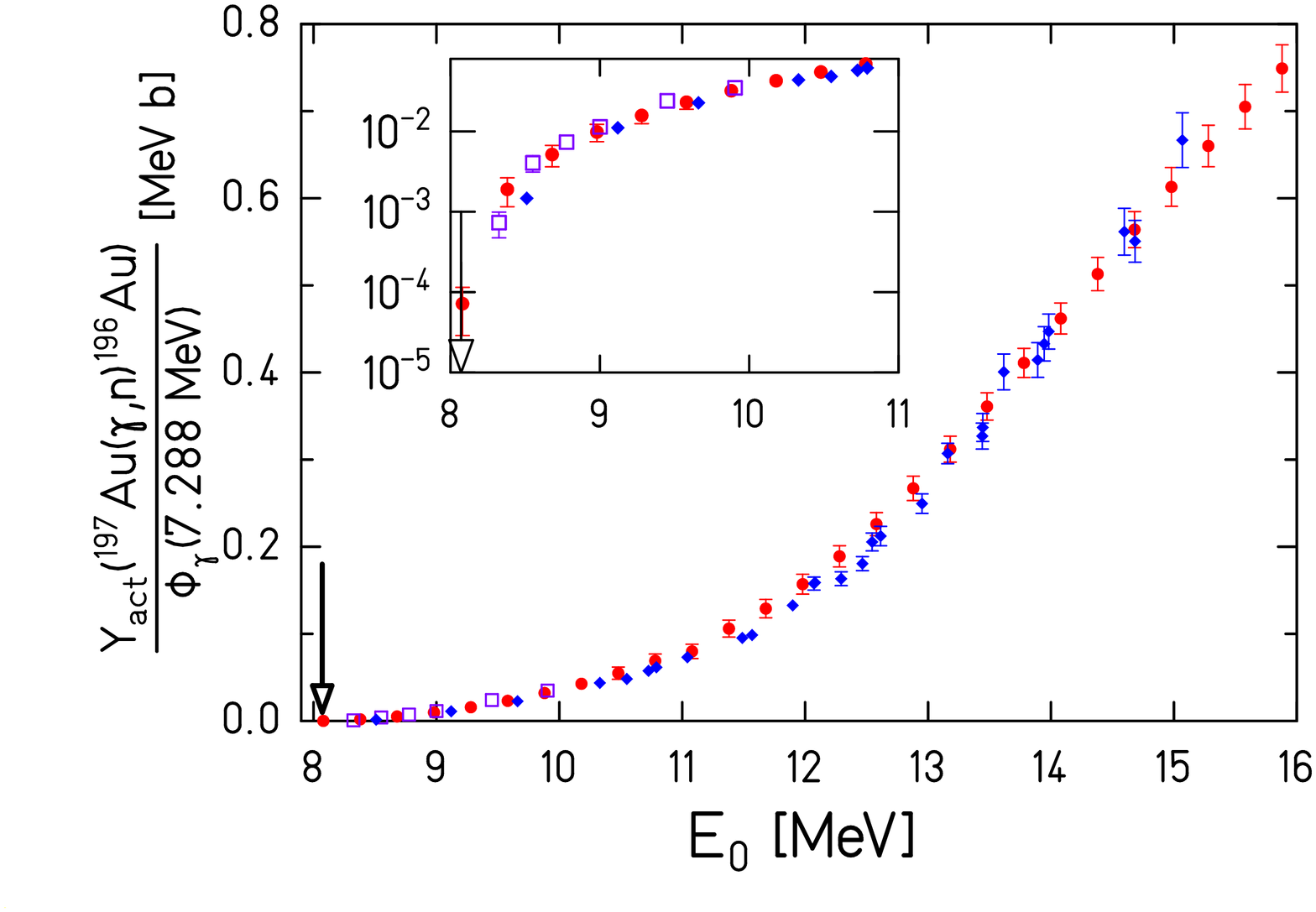}
\end{center}
\caption{\label{fig:yield-exptvsexpt}(Color online) Activation yield
for the $^{197}$Au$(\gamma,n)$ reaction normalized to the photon
fluence is compared to the yield calculated using cross sections
measured in previous experiments. The present data are denoted by
diamonds with an arrow pointing to the neutron emission threshold.
Reaction yield calculated using the cross sections given by
Veyssiere et al.~\cite{Veyssiere1970} (circles) and Vogt et
al.~\cite{Vogt2002} (open squares) are in good agreement with the
yield measured at ELBE.}
\end{figure}

In Fig.~\ref{fig:yield-exptvsexpt}, the experimental activation
yield from ELBE is compared to the yield calculated using the cross
sections measured previously. The activation yield is normalized to
the photon fluence measured from the scattered photons in $^{11}$B
(see Sect.~\ref{sect:flux}). The experimental yield from ELBE is in
agreement with the yield calculated using the cross sections from
Vogt et al. for the close-threshold endpoint-energies up to 10 MeV.
The activation yield calculated using cross sections from Veyssiere
et al. is in agreement with the ELBE yield for the whole range of
energies. Close to the neutron emission threshold, the reaction
yield strongly depends on the endpoint energy $E_0$ of the
bremsstrahlung beam. In this case, small uncertainties in $E_0$
result in large uncertainties of the activation yield.

The uncertainties in the experimental points shown in
Fig.~\ref{fig:yield-exptvsexpt} are mainly from the determination of
photon fluence as discussed in Sec.~\ref{sect:flux}. The statistical
uncertainties are very small and in the order of about 0.5-2\%. The
major systematic uncertainties arise from the extrapolation of
measured photopeak efficiencies to the higher energies in $^{11}$B
transitions (5\%) and in the systematic deviation of measured photon
fluence from the simulated curve (6\%). The systematic errors have
been added quadratically and amount to about 7.8\% but are not shown
in Fig.~\ref{fig:yield-exptvsexpt}.

\subsection{Activation Yield : Comparison with model
calculations}\label{sect:statisticalmodel}
\begin{figure}
\begin{center}
\includegraphics[width=6.2 cm,angle=270]{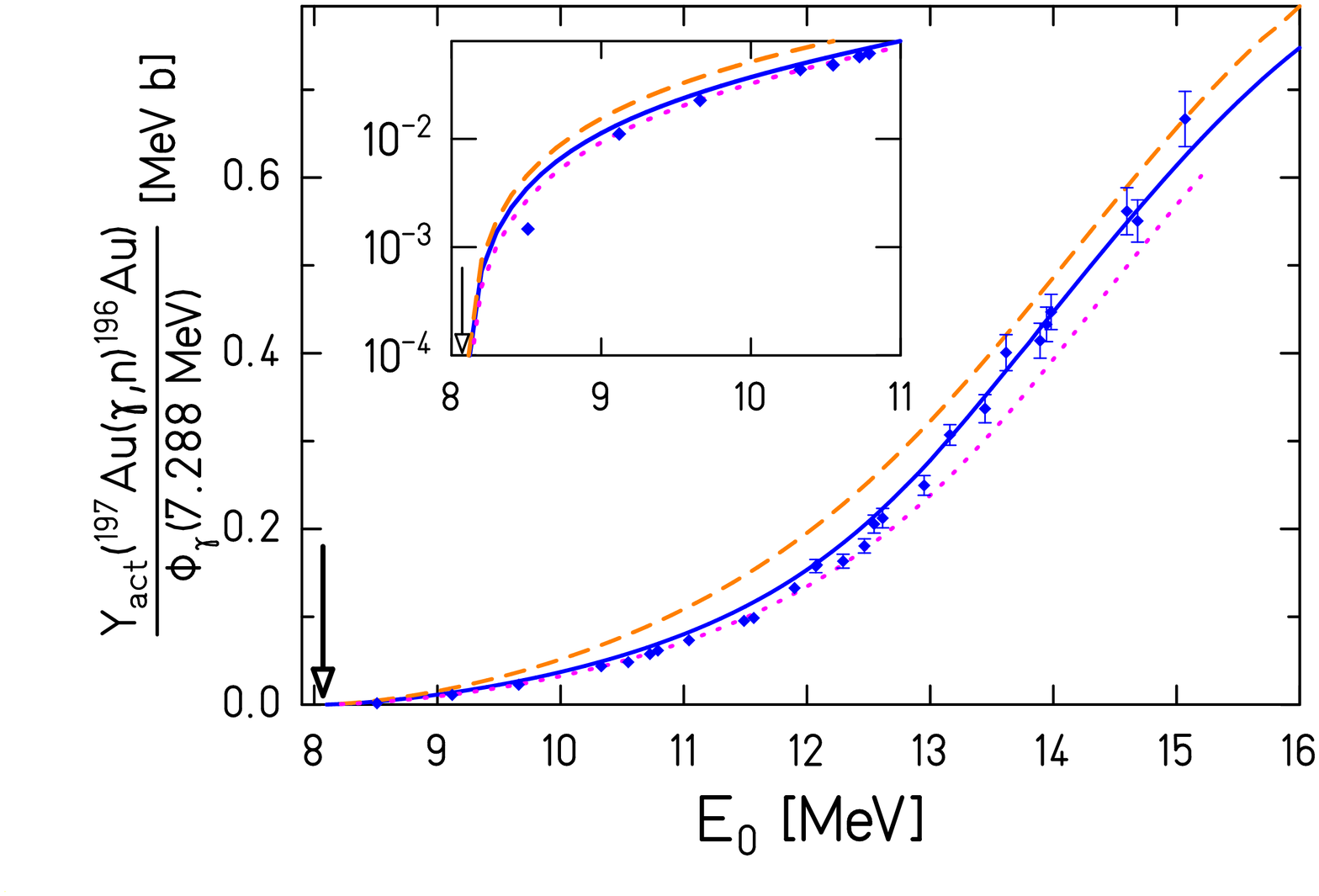}
\end{center}
\caption{\label{fig:yield-exptvstheory}(Color online) Experimental
activation yield normalized to the photon fluence for the
$^{197}$Au$(\gamma,n)$ reaction compared to theoretical model
calculations. The experimental data are denoted by diamonds with a
downward arrow denoting the neutron emission threshold. The dashed
and dotted lines denote yield calculations using cross sections from
the TALYS~\cite{Koning2005} and NON-SMOKER~\cite{Rauscher2004} codes
respectively. The solid line represents a TALYS calculation with
modified inputs, see text.}
\end{figure}

Fig.~\ref{fig:yield-exptvstheory} compares the experimental
activation yield to the simulated yield calculated using cross
sections predicted by Hauser-Feshbach
models~\cite{Koning2005,Rauscher2004}. Simulations using the
TALYS~\cite{Koning2005} and NON-SMOKER~\cite{Rauscher2004} codes
describe the experimental data only to a factor of 2. Both
calculations were performed using cross sections derived from
standard input parameters. The default option of TALYS for the GDR
parameters originates from the Beijing GDR compilation, as present
in the RIPL database~\cite{RIPL-2}.

In the case of $(\gamma,n)$ reactions, one crucial ingredient for
the model calculation is the photon strength function. As the
$(\gamma,n)$ channel in $^{197}$Au is the dominant decay channel for
the energy range above threshold, the photon strength distribution
directly determines the calculated $(\gamma,n)$ cross section and
the reaction yield. In the model calculations care is also taken for
the fact that the $(\gamma,p)$ channel is open above 5.8 MeV. Due to
the large $Z$, the $p$-emission is strongly suppressed by the
Coulomb interaction. With default inputs, the TALYS calculation
yields a $(\gamma,p)$ cross section which is about four orders of
magnitude smaller than the $(\gamma,n)$ cross section.

The activation yields calculated using TALYS with different optical
model potentials, like Koning-Delaroche and Jeukenne-Lejeune-Mahaux
(JLM), are very similar demonstrating that the
$^{197}$Au$(\gamma,n)$ reaction yield is not sensitive to the choice
of optical model parameters. The sensitivity to the  photon strength
function is larger. We modified the deformation dependent parameters
of the E1 strength function used in TALYS according to a new
phenomenological parametrization. The improved new parametrization
explains the experimental data better than the statistical models
with default inputs and is discussed in the following section in
detail.

\subsection{Phenomenological parametrization of the photon strength function}\label{sect:parametrization}
If one assumes that the dipole strength in a heavy nucleus is
dominated by the GDR, then the strength function $f_{1}(E_{\gamma})$
according to Bartholomew et al.~\cite{Bartholomew1973} is related to
the average photoabsorption cross section
$\left<\sigma_{\gamma}(E_{\gamma})\right>$ by
\begin{equation}
\frac{\left<\sigma_{\gamma}(E_{\gamma})\right>}{3(\pi\hbar\mathrm{c})^2\cdot
E_{\gamma}} = f_{1}(E_{\gamma}) =
\frac{\left<\Gamma_{E_{1}}\right>}{E_{\gamma}^3\cdot D},
\label{eq:sigma}
\end{equation}
with $\left<\Gamma_{E_{1}}\right>$ and $D$ denoting the average
photon width and the average level spacing at the endpoint of
electromagnetic transition. A new phenomenological description based
on the ground state deformation parameters describes well the
average photon absorption for nuclei with $A$ $>$ 80 from
$E_{x}\approx$ 4 MeV up to several MeV above the
GDR~\cite{Junghans2008}.

A consistent description holds for the photon strength distribution
in spherical, transitional, triaxial and well deformed nuclei. In
nearly all nuclei the GDR is split into two or three components,
whose energies are well predicted by the finite range droplet model
(FRDM)~\cite{Myers1977}. The splitting~\cite{Bush1991} is due to the
three different axes of the ellipsoid parameterizing the nuclear
shape with its deformation parameter $\beta$ and triaxiality
parameter $\gamma$:
\begin{equation}
E_{k} = \frac{E_{0}\cdot R_{0}}{R_{k}} =
\frac{E_{0}}{\exp\left[\sqrt{\frac{5}{4 \pi}}\cdot \beta
\cdot\mathrm{cos}(\gamma-\frac{2}{3}k\pi)\right]} \label{eq:GDR}
\end{equation}
This results from the fact that the vibrational frequency
$E_{k}/\hbar$ along a given axis $k$ is inversely proportional to
the corresponding semi-axis length $R_{k}$. The nuclear radius is
given by $R_{0}=1.16 A^{1/3}$fm. The GDR centroid energy $E_{0}$
given in Ref.~\cite{Myers1977} of a spherical nucleus with mass $A$
is calculated with an effective nucleon mass $m^*$= 874 MeV/c$^2$.

The average absorption cross section in the GDR is given by
\begin{equation}
\left<\sigma_{\gamma}(E_{\gamma})\right> = \frac{1.29\cdot Z\cdot
N}{A}\sum_{\rm
k=1}^3\frac{E_{\gamma}^2\Gamma_{k}}{(E_{k}^2-E_{\gamma}^2)^2+E_{\gamma}^2\Gamma_{k}^2}
\label{eq:Gamma}
\end{equation}
where the GDR widths $\Gamma_{k}$ to be used in the sum of up to
three Lorentzians have been assumed to be constant, in contrast to
earlier descriptions~\cite{Kopecky1941,Zanini2003}. The symbols
$E_\gamma$, $E_k$ denote photon energy and resonance energies given
in MeV and $\left<\sigma_{\gamma}(E_{\gamma})\right>$ given in
fm$^2$. The Thomas-Reiche-Kuhn sum rule as determined from general
quantum mechanical arguments~\cite{Eisenberg1988} is included in
this description for the average photon absorption cross section
obtained on an absolute scale.

The width $\Gamma_{k}$ for the different components of the GDR is
dependent on the resonance energy $E_k$ and is generally used for all stable
nuclei with $A>$ 80
\begin{equation}
\Gamma_{k}(E_{k}) = 1.99 \mbox{ MeV}\cdot
\left(\frac{E_{k}}{10 \mbox{ MeV}}\right)^\delta ,
\label{eq:gammak}
\end{equation}
where $\delta$ = 1.6 is taken from the one body dissipation model~\cite{Bush1991}.

For the case of $^{197}$Au we assume that the average of the
experimentally determined deformation parameters of the even-mass
neighbor nuclei $^{196}$Pt and
$^{198}$Hg~\cite{Mauthofer1990,Bockisch1979,Stone2005} can be used
to describe the shape of the odd nucleus $^{197}$Au, we insert
$\beta = 0.15$ and $\gamma$ = 60$^{\circ}$ into Eq.~(\ref{eq:GDR}).
The GDR centroid energy is $E_{0}$= 13.9 MeV. These parameters are
in accordance with the FRDM and result in the following resonance
energies and widths: $E_{1,3}$= 13.2 MeV, $\Gamma_{1,3}$ = 3.1 MeV
and $E_{2}$= 15.2 MeV, $\Gamma_{2}$ = 3.9 MeV.  The TALYS code was
modified with these inputs for oblate deformation. The yield curve
created using the cross sections resulting from modified inputs is
shown in Fig.~\ref{fig:yield-exptvstheory} and is in better
agreement to the ELBE data.

\begin{figure}
\begin{center}
\includegraphics[trim= 10 50 10 0, width=6 cm,angle=270]{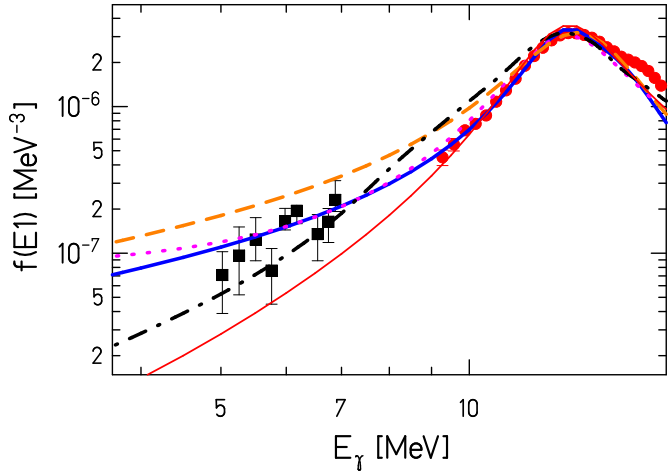}
\end{center}
\caption{\label{fig:strength-comparison}(Color online) The photon
strength function of $^{197}$Au derived on the assumption of oblate
deformation (solid line) compared to different models. The dashed
and dotted lines correspond to the strength functions given by
Brink-Axel~\cite{Brink-Axel} and Kopecky-Uhl~\cite{Kopecky1941}
models respectively. The microscopic E1 photoabsorption
strength-functions determined within the QRPA
model~\cite{Goriely2002,Khan2001} is shown by the dash-dotted line.
All calculations were done using the TALYS Code. The Enhanced
Generalized LOrentzian (EGLO) model taken from the Reference Input
Parameter Library RIPL-2 of the IAEA~\cite{RIPL-2} is shown as a
thin solid line. The experimental strength function from Bartholomew
et al.~\cite{Bartholomew1973} (squares) below the neutron emission
threshold and the strength function derived using the
$^{197}$Au($\gamma$,n) photoneutron cross section measured by
Veyssiere et al. (circles) are also shown.}
\end{figure}

The  photon strength function of $^{197}$Au derived from different
theoretical models and compared to experimental data is shown in
Fig.~\ref{fig:strength-comparison}. The strength function created
using the modified inputs as discussed above is compared to the
default models~\cite{Brink-Axel,Kopecky1941} in TALYS which treats
$^{197}$Au as a spherical nucleus. It is clear that the new
parameters lead to a reduced strength at energies below the GDR and
thus result in a good fit to its shape with a constant spreading
width. This agrees well to the experimental strength function given
by Bartholomew et al.~\cite{Bartholomew1973} for energies below the
neutron emission threshold. Above the separation energy, the
strength functions shown were deduced from the
$^{197}$Au$(\gamma,n)$ cross sections by Veyssiere et al. The
strength function derived using the modified parameters gives
clearly a better fit to the data than
calculations~\cite{Goriely2002,Khan2001} on the basis of the quasi
particle random phase approximation (QRPA) with phenomenological
correction for deformation.

\section{Conclusions}
For the $^{197}$Au$(\gamma,n)$ reaction, the activation yield has
been measured and compared to the Hauser-Feshbach model calculations
as well as previous experimental data. The measured activation yield
at ELBE is in agreement to the calculated yields using cross
sections measured with quasi-monoenergetic photons from positron
annihilation in flight and laser-induced Compton backscattering.

The activation experiment discussed here, which was performed in
combination with a direct determination of the electron energy via
the bremsstrahlung spectrum endpoint, deliver precise photon
strength data. Thus they may allow a verification of data obtained
previously by direct absorption experiments or by detecting neutrons
from the $(\gamma,n)$-process. They also allow us to make judgements
on parameterizations developed for the prediction of the photon
strength function as well as on particle transmission functions,
i.e. on optical model parameters.

We have demonstrated for the case of $^{197}$Au$(\gamma,n)$ that a
sum of two GDR-Lorentzians with a small oblate deformation of
$^{197}$Au determining the energy split and the width difference
describes the photon absorption well. The availability of
information on the nuclear shape as well as on the PSF below
threshold makes $^{197}$Au a prime case to perform a consistent test
of statistical calculations of Hauser-Feshbach type and to derive a
coherent picture of near-threshold processes. The detailed
understanding of these has direct importance for the s-process as
well, since the prediction of the relative abundances of the
isotopes of Hg depends on the relative strength of the $\beta$-decay
of $^{198}$Au and $^{198}$Au$(n,\gamma)$ in stellar plasmas. Last,
but not least, the experimental data reported here for the
$^{197}$Au$(\gamma,n)$ reaction may serve as a normalization for
future measurements on other nuclei. The $^{197}$Au$(\gamma,n)$
reaction has been used as a photoactivation standard for the
experiments discussed in Refs.~\cite{Nair2007,ErhardEPJ}.

\section{Acknowledgements}
We thank Peter Michel and the ELBE team for providing a stable beam
during activation experiments. We are indebted to E. Haug for
providing valuable information about the theory of bremsstrahlung
and his model codes. Special thanks to Rudi Apolle for simulating
proton spectra of the deuteron breakup reaction under a realistic
geometry. The technical assistance of Andreas Hartmann is gratefully
acknowledged. We are grateful to Tom Cowan for a thorough, critical
reading of the manuscript.

\end{document}